\documentstyle[aps,epsfig,amsmath]{revtex} 
\newcommand{\qs}{\tilde{q}}
\newcommand{\erf}{{\rm erf}}

\title{Replica symmetry breaking solution for the fermionic Ising spin glass and the Ghatak--Sherrington model}
\author{H. Feldmann and R. Oppermann}
\date{}

\begin{document}
\maketitle
\begin{abstract}
  We solve the fermionic version of the Ising spin glass for arbitrary
  filling $\mu$ and temperature $T$ taking into account replica
  symmetry breaking. Using a simple exact mapping from $\mu$ to the
  anisotropy parameter $D$, we also obtain the solution of the $S=1$
  Sherrington--Kirkpatrick model.  An analytic expression for $T=0$
  gives an improved critical value for the first--order phase
  transition.  We revisit the question of stability against
  replica--diagonal fluctuations and find that the appearance of
  complex eigenvalues of the Almeida--Thouless matrix is not an
  artifact of the replica--symmetric approximation.
\end{abstract}

\section{Introduction and Mapping of the two models}
A direct generalization of the Sherrington--Kirkpatrick (SK) model
\cite{SK} to include quantum fluctuations are fermionic spin glasses
\cite{oppermann93a}. They provide a larger class of models, since the
partition functions of all classical spin glass models can be extracted
from them through the Popov--Fedotov trick \cite{popov88a,oppermann94a}.
They also allow to extend standard spin glass theory
to itinerant systems, to study the influence of spin glass order on
the excitation spectrum \cite{oppermann98a} and to investigate the
competition of spin glass order with other kinds of ordering typical
of quantum systems \cite{oppermann99a}. All these aspects may be
relevant to the physics of heavy fermion compounds.

Another extension of the SK model is the $S=1$ spin glass in a crystal
field, realized for example in $\rm (Ti_{1-x}V_x)_2O_3$ \cite{ghatak77a}. Both
extensions show tricritical behavior as the chemical potential $\mu$
(for fermionic spin glasses) or the anisotropy parameter $D$ are
varied \cite{ghatak77a,oppermann96a}. A simple mapping relates the two
models, as far as the static properties are concerned.

The fermionic Ising spin glass (ISG$_f$) is described by the grand
canonical Hamiltonian
\begin{equation}
{\cal H} = - \sum_{ij}J_{ij} \sigma_i \sigma_j - \mu \sum_i n_i
\label{eq:hamiltonianISGf}
\end{equation}
The coupling $J_{ij}$ is Gaussian distributed around 0 with variance
$J^2$. Below, we always set $J=1$.  The main difference to the SK spin
glass is that spins and occupation numbers are given in terms of
fermionic operators which act on a space with four states per site,
$|00\rangle$, $|\uparrow0\rangle$, $|0\downarrow\rangle$, and
$|\uparrow\downarrow\rangle$:
\begin{equation}
\sigma = a_{\uparrow}^{\dagger}a_{\uparrow} 
- a_{\downarrow}^{\dagger}a_{\downarrow} \qquad
n=a_{\uparrow}^{\dagger}a_{\uparrow} 
+ a_{\downarrow}^{\dagger}a_{\downarrow}
\end{equation}
To obtain the thermodynamic behavior of the model, we calculate the
free energy using the replica trick \cite{parisi87a}. Integrating over
the distribution of $J_{ij}$ creates eight--fermion correlations,
which are decoupled using the Parisi matrix of order parameters
$Q^{a\tau,b\tau'}$. For mean field theory, we use the static saddle
point of this matrix.  Details of this calculation can be found in
\cite{oppermann93a,oppermann98b}.

The $S=1$ anisotropic spin glass or Ghatak--Sherrington (GS) model
\cite{ghatak77a} is represented by the Hamiltonian
\begin{equation}
{\cal H} = -  \sum_{ij} J_{ij} S_{iz} S_{jz} + D \sum_i S_{iz}^2
\label{eq:hamiltonianGS}
\end{equation}
where $S_z$ may have the values $-1$,$0$, and $1$.  It is easy to see
that the partition function corresponding to the Hamiltonian
(\ref{eq:hamiltonianISGf}) is identical --- apart from the constant $1
+ \exp(- 2 \beta \mu)$ --- to the one defined by Eq.
(\ref{eq:hamiltonianGS}), provided one maps anisotropy and chemical
potential according to
\begin{equation}
e^{\beta D}=e^{\beta \mu} + e^{-\beta \mu}
\label{eq:mapping}
\end{equation}
As a consequence of (\ref{eq:mapping}), the thermodynamic properties
of both models are directly related. At $T \to 0$, one may even
neglect the last term and set $D=\mu$. Note, however, that there is a
whole class of fermionic correlations that can not be expressed in
terms of $S_z$ and are therefore unique to the ISG$_f$, because their
definition requires a fermionic generating functional. A good example
is the fermion Green's function \cite{oppermann98d}. Also, the
Heisenberg and XY variant of the ISG$_f$, since they exhibit quantum
dynamics, do not have a direct classical analogue.

In passing, we note that there is another interesting connection
between the two models in addition to the mapping (\ref{eq:mapping}) and the
above mentioned Popov--Fedotov method. One could introduce a repulsive 
Hubbard--like on--site interaction $U$ to Hamiltonian 
(\ref{eq:hamiltonianISGf}), which would deplete the doubly occupied 
states in the limit of infinite strength. In this case, the Hamiltonian
and a mapping similar to (\ref{eq:mapping}) would read
\begin{equation}
{\cal H} = - \sum_{ij} J_{ij} \sigma_i \sigma_j - (\mu-\frac{U}{2})\sum_i n_i
+ U \sum_i (a_{\uparrow}^{\dagger}a_{\uparrow}-\frac12) 
(a_{\downarrow}^{\dagger}a_{\downarrow}-\frac12) \quad \mbox{and} \quad
e^{\beta D} = e^{-\beta \mu} + e^{\beta \mu - \beta U}
\end{equation}

\section{Results at nonzero temperature}
\label{sec:nonzeroT}
It has been known for a long time that the two models under
consideration are described by two mean--field parameters, $q(x)$ and
the replica--diagonal saddle--point value, $Q^{aa}=\tilde{q}$ (in the
notation of the ISG$_f$).

Already from replica--symmetric (RS) calculations, one can locate the
second--order critical line, terminating at the tricritical point at
($T=1/3$, $D=0.962$ or $\mu=0.961$). In order to correctly describe the system
below this line, one has to allow for replica symmetry breaking (RSB).
Since only few results are available for full RSB \cite{oppermann99a},
the first step of the Parisi scheme (1RSB) \cite{parisi80a} can give
valuable information. In many cases, it already comes close to the
full solution or allows a good conjecture.

Using the standard form of RSB at the one--step level, we use the
order parameters $q_1=q(1)$ and $q_2=q(0)$. The position of the jump
in $q$ is given by $m$.  After integration over the fermionic fields
and some other transformations we obtain the replica--broken free
energy

\begin{align}
  \beta f =&\frac{\beta^{2} J^{2}}{4} ((\qs-1)^{2} - (q_1-1)^{2} +
  m(q_1^2 - q_2^2)) - \beta \mu - \frac1m \int_{z_2}^{G} \ln
  \int_{z_1}^G{\cal{C}}^m
  \quad \mbox{with}\label{eq:f1RSB}\\
  {\cal{C}} =& \cosh(\beta h + \beta J \sqrt{q_1-q_2} z_1 +\beta J
  \sqrt{q_2} z_2) + \cosh (\beta \mu) e^{- \frac{\beta^{2}
      J^{2}}{2}(\qs - q_1)}
\end{align}

\begin{figure}
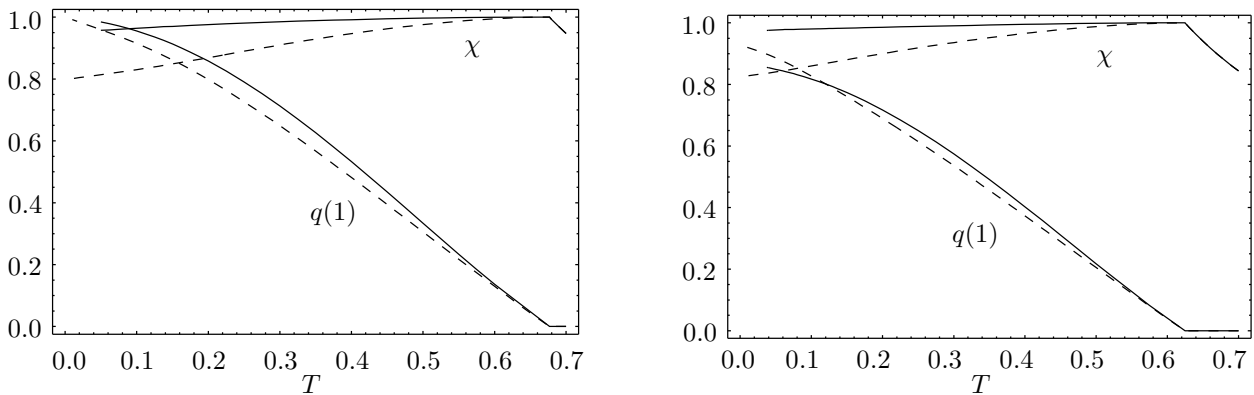

  \input{order_mu_0.0.pstex_t} \hspace{1cm}
  \input{order_mu_0.5.pstex_t} \vspace{0.3cm}
\caption{Susceptibility $\chi$ and Edwards-Anderson order parameter $q(1)$ for the ISG$_f$ for half--filling and for $\mu=0.5$. Solid and dashed lines show the results from the 1--step replica broken and the replica--symmetric solutions, respectively.}
\label{fig:finiteTOrderParameters}
\end{figure}

The mean--field solutions of the model are given by the condition that
$f$ is stationary with respect to $\qs$, $q_1$, $q_2$, and $m$.
Considerable numerical effort is necessary to solve the resulting
self--consistency equations for these four parameters simultaneously,
and for all values of chemical potential and temperature within the
ordered phase. In Fig.\ref{fig:finiteTOrderParameters} we present a
selection of our results, introducing the susceptibility
$\chi=\beta(\qs-(1-m)q_1-m \, q_2)$. We compare our results with
results from the replica--symmetric calculation. Note that $\chi$ in
1RSB is already very close to the expected exact value $\chi=1$ in the
ordered phase, especially for temperatures close to the phase
transition.

\section{Zero temperature results}
\label{sec:zeroT}
In the limit of vanishing temperature, $q=\qs + O(T)$. Also $m$
vanishes as $T \to 0$. Therefore, for the zero--temperature behavior
of the model, we have to use a new set of parameters in order to avoid
divergences in Eq.  (\ref{eq:f1RSB}).  We replace $\beta J (\qs-q_1)$
by the single--valley susceptibility $\overline{\chi}$ and $m$ by $ a
T$. Note that although the $m$, the size of the ``lower step'' in
$q(x)$, vanishes with the temperature, replica symmetry breaking still
has a profound effect on the system.

In these new variables, the limit of Eq. (\ref{eq:f1RSB}) can be
written as
\begin{equation}
  f=\frac12 J \overline{\chi} (q_1 -1)+ \frac14 J a (q_1^2-q_2^2)
  - \mu -\frac{J}{a}\int_{z_2}^G \ln I
\label{eq:f1RSB0T}  
\end{equation}
where $I$ is given for the case $\mu/J<\overline{\chi}/2$ by
\begin{equation}
  \begin{split}
    I=& \frac12 e^{a \sqrt{q_2}z_2+ \frac12 a^2(q_1-q_2)}
    \Bigl(1+\erf\bigl(\frac{\sqrt{q_2}z_2 + a (q_1-q_2)}{\sqrt{q_1-q_2}\sqrt{2}}\bigr)\Bigr)\\
    &+\frac12 e^{-a \sqrt{q_2}z_2+ \frac12 a^2(q_1-q_2)}
    \Bigl(1+\erf\bigl(\frac{-\sqrt{q_2}z_2 + a
      (q_1-q_2)}{\sqrt{q_1-q_2}\sqrt{2}}\bigr)\Bigr)
  \end{split}
\label{eq:int1case1}
\end{equation}
and for the case $\mu/J>\overline{\chi}/2$ by
\begin{equation}
\begin{split}
  I =& \frac12 e^{a \sqrt{q_2}z_2 + \frac12 a^2(q_1-q_2)}
  \Bigl(1+\erf\bigl(\frac{\sqrt{q_2}z_2 - (\frac{\mu}{J}-\frac{\overline{\chi}}{2})+a (q_1-q_2)}{\sqrt{q_1-q_2}\sqrt{2}}\bigr)\Bigr)\\
  +& \frac12 e^{a(\frac{\mu}{J}-\frac{\overline{\chi}}{2})} \Bigl(
  \erf\bigl(\frac{(\frac{\mu}{J}-\frac{\overline{\chi}}{2})-\sqrt{q_2}z_2}{\sqrt{q_1-q_2}\sqrt{2}}\bigr)
  +
  \erf\bigl(\frac{(\frac{\mu}{J}-\frac{\overline{\chi}}{2})+\sqrt{q_2}z_2}{\sqrt{q_1-q_2}\sqrt{2}}\bigr) \Bigr)\\
  +& \frac12 e^{-a \sqrt{q_2}z_2 + \frac12 a^2(q_1-q_2)} \Bigl(1+
  \erf\bigl(\frac{-\sqrt{q_2}z_2-(\frac{\mu}{J}-\frac{\overline{\chi}}{2})+a(q_1-q_2)}{\sqrt{q_1-q_2}\sqrt{2}}\bigr)\Bigr)
\end{split}
\label{eq:int1case2}
\end{equation}

\begin{figure}
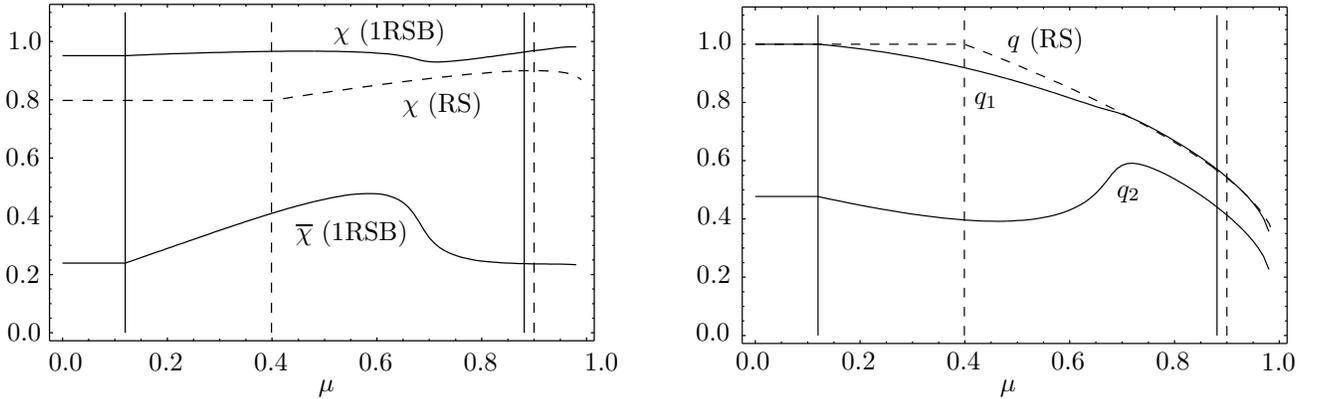

  \input{zeroTsusceptibility.pstex_t} \hspace{1cm}
  \input{zeroTOrderParameters.pstex_t} \vspace{0.3cm}
\caption{Normal ($\chi$) and single--valley ($\overline{\chi}$) susceptibility (left) and spin--glass order 
  parameters (right) at zero temperature. Solid curves correspond to
  the replica broken solution, dashed curves to the replica--symmetric
  one. For each case, the left vertical line indicates the onset of
  the $\mu$--independent solution for small $\mu$, while the right
  vertical line locates the first order transition to the paramagnetic
  regime.}
\label{fig:zeroTOrderParameters}
\end{figure}

Again, variation with respect to the order parameters gives four
equations describing the saddle point solution. We solved these for
all chemical potentials within the ordered phase and present the
results for $\qs=q_1$, $\chi$ and $\overline{\chi}$ in Fig.
\ref{fig:zeroTOrderParameters}, together with the replica symmetric
order parameters for comparison. In RS approximation,
$\overline{\chi}=\chi$, which illustrates the importance of RSB.

We also calculated the free energy from Eqs.
(\ref{eq:f1RSB0T})-(\ref{eq:int1case2}) and compared it with the free
energy for the paramagnetic phase, $f_{\rm para}=-2 \mu$, to obtain
the first order phase transition at $\mu_{\rm t,1}=0.881$, a slightly
lower value compared with the replica--symmetric approximation, which
gave $\mu_{\rm t,0}=0.900$.

The transition between the two regimes described by Eqs.
(\ref{eq:int1case1}) and (\ref{eq:int1case2}) takes place at
$\mu=0.119$, which is also related to the width of the band gap at
half filling \cite{oppermann98d,oppermann98b}.

\section{Replica symmetry breaking in charge correlations}

As sections \ref{sec:nonzeroT} and \ref{sec:zeroT} show, RSB has a
profound effect on the low temperature phase. However, spin and
charge correlations are affected differently. While spin correlations
are modified by RSB within the whole ordered phase (see the
susceptibility in Figs. \ref{fig:finiteTOrderParameters} and
\ref{fig:zeroTOrderParameters}), charge correlations remain
essentially unchanged by RSB within part of the phase diagram.

\begin{figure}
  \centerline{\input{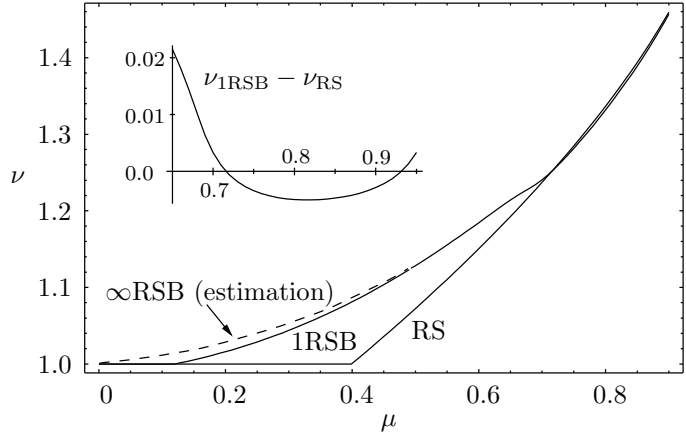}} \vspace{0.3cm}
\caption{$\mu$ -- dependence of the filling at zero temperature in RS and 1RSB approximation. For $\mu<0.119$, both approximations yield a constant $\nu$, while the full solution, estimated with the dashed line, gives $\nu=1$ only at $\mu=0$. Note the remarkable drop of the RSB solution to the RS line around $\mu=0.7$. After the crossing the two lines stay close together. The inset shows the difference between the 1RSB and the RS solutions for large $\mu$.}
\label{fig:fillingZeroT}
\end{figure}

The most obvious observable parameter connected to the
replica--diagonal order parameter $\qs$ is the filling $\nu$. Both are
related to all orders of RSB by
\begin{equation}
\qs = 1 - \coth(\beta \mu)(\nu-1)
\end{equation}
Zero--temperature results for $\nu$ are displayed in Fig.
\ref{fig:fillingZeroT}. The difference between the broken and the
unbroken solution is largest at the point where the RS value of $\nu$
turns constant. Around $\mu=0.7$, the 1RSB solution bends around to
come close to the RS one and the two solutions cross. But the
difference stays comparatively small, so for large $\mu$ the filling
is virtually not affected by RSB. A similar situation appears in the
standard spin glass theory for $m>1$ component spin glasses. There, in
a strong magnetic field two characteristic lines appear, first the
Gabay--Toulouse line \cite{toulouse81a,moore82a,sherrington82a}, where
RSB appears for the order parameter perpendicular to the field while
it has little effect on the longitudinal order parameter. At a lower
temperature, close to the line found by de Almeida and Thouless, also
the longitudinal order parameter strongly feels RSB. In this context,
one may view the charge and spin degrees of freedom to be
``orthogonal'' and the chemical potential to correspond to a magnetic
field.

\begin{figure}
  \centerline{\input{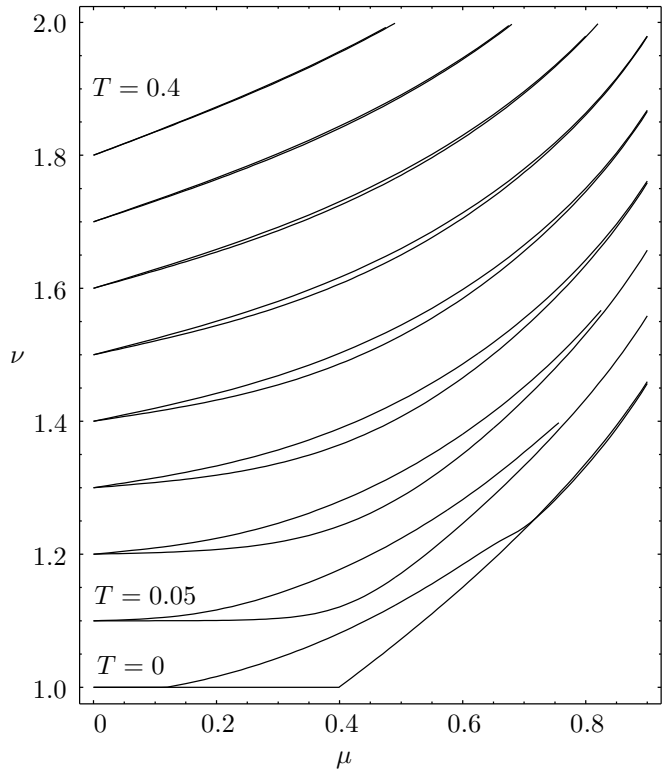}} \vspace{0.3cm}
\caption{Filling factor $\nu$ as a function of the chemical potential 
  for temperatures $T=0,0.05,0.1 \dots 0.4$. For each pair of curves,
  the lower one gives the RS solution, the upper one the 1RSB result.
  We applied an offset in steps of 0.1 to separate the results for
  different $T$. For $\mu=0$, there is always half--filling.  Note
  that the effect of RSB on $\nu(\mu)$ is significant only in a
  intermediate range of chemical potentials and for small
  temperatures.}
\label{fig:filling}
\end{figure}

For nonzero temperatures, the difference between RS and 1RSB in
general is smaller than for $T=0$, as shown in Fig. \ref{fig:filling}.
The merging of the two lines becomes smoother with increasing
temperature.  The effect of broken replica symmetry on $\nu$ already
is invisible at $T=0.4$ in Fig. \ref{fig:filling}, far below the
second order phase transition (see Fig. \ref{fig:stability0rsb}).

For the GS model, the results of this section on the filling can be
directly translated to the average number of sites with $S_z=0$:
\begin{equation}
\qs = \langle S_z^2 \rangle \quad \mbox{or} \quad
\langle 1- S_z^2 \rangle = \coth(\beta \mu)(\nu-1)
\end{equation}

\section{Replica--diagonal stability}

The replica--symmetric solution is unstable against replica--symmetry
breaking, as shown by de Almeida and Thouless (AT) in \cite{AT}.  This
problem, together with its famous solution using an ultrametric
saddle--point matrix \cite{parisi87a}, is one of the main reasons for
the enormous theoretical interest in spin glasses over the past
decades.

In addition to the eigenvalue that marks the onset of
replica--symmetry breaking, AT obtained two additional pairs of
eigenvalues which merge in the replica limit. For the
Sherrington--Kirkpatrick model, these were positive for all
temperatures and did not pose a problem. In the case of the
Ghatak--Sherrington model and the ISG$_f$, however, the
replica--diagonal eigenvalues can become complex, as noted by several
authors \cite{lage82a,mottishaw85a,Dacosta94a,oppermann98b}.

This result is very difficult to interpret. Lage and de Almeida
\cite{lage82a} derived another set of stability conditions and found
their condition $\partial^2f/\partial \qs^2$ to be violated at low
temperatures, in a region determined by the ``crossover line'' of Ref.
\cite{oppermann99a}.  Mottishaw and Sherrington \cite{mottishaw85a}
pointed out that the system is unstable against RSB and they suspected
that the full Parisi solution will have only real eigenvalues again.

\begin{figure}
  \centerline{\input{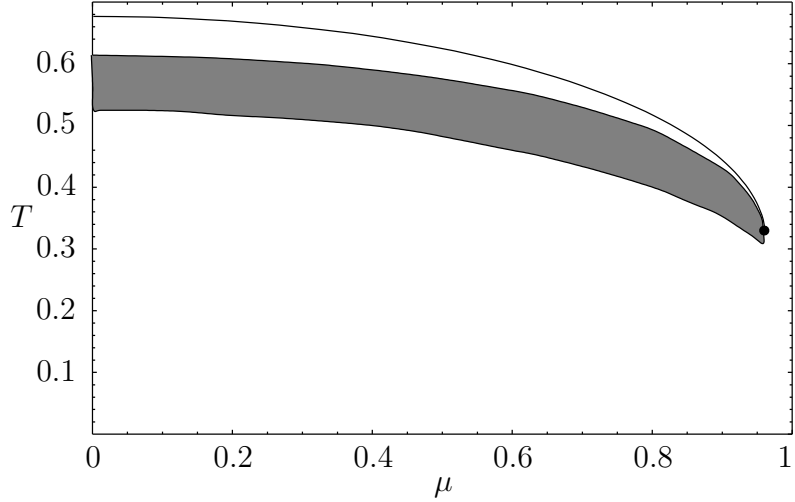}} \vspace{0.3cm}
\caption{Region of complex replica--diagonal AT eigenvalues in the
  replica--symmetric approximation. Also shown is the line of
  continuous transitions ending in the tricritcal point.}
\label{fig:stability0rsb}
\end{figure}

We have numerically evaluated the self--consistency equations from
Ref. \cite{oppermann93a} within the ordered phase and calculated the
replica--diagonal eigenvalues. Directly below the second--order phase
transition and at low temperatures, they are real and positive,
indicating stability against replica--diagonal fluctuations. But in
between there is a region extending from $\mu=0$ to the tricritical
point, where they have an imaginary part, while the real part remains
positive (see Fig.  \ref{fig:stability0rsb}).  At half--filling, the
complex eigenvalues occur between $T=0.53$ and $T=0.61$.

To extend the AT analysis of the replica--diagonal eigenvalues to
replica symmetry breaking, we start with the ansatz for the
eigenvector
\begin{equation}
\mu =\binom{\{\epsilon^{(aa)}\}}{\{\eta^{(a b)}\}}
\end{equation}
where the replica indices run from $a,b=1$ to $n$ and only pairs with
$a<b$ are considered for $\eta^{(a b)}$.  For $k$ step RSB, the
replica--diagonal eigenvector of AT can be generalized to
\begin{equation}
\epsilon^{(aa)}= \alpha \quad \mbox{and} \quad
\eta^{(ab)}=\begin{cases}
\beta_1 & \mbox{if}\;\; Q^{ab}=q_1 \\ 
\beta_2 & \mbox{if}\;\; Q^{ab}=q_2 \\
\dots & \\ 
\beta_{k+1} & \mbox{if}\;\; Q^{ab}=q_{k+1} \\
\end{cases}
\label{eq:vectorAnsatz}
\end{equation}

Substituting Eq. (\ref{eq:vectorAnsatz}) into the quadratic form
\begin{equation}
\mu^T G  \mu \quad \mbox{with} \quad G^{(ab)(cd)}=
\frac{\partial}{\partial Q^{ab}}
\frac{\partial}{\partial Q^{cd}} f
\end{equation}
yields $k+2$ replica--diagonal eigenvalues.  In the limit of $n \to
0$, these eigenvalues can also be represented as the eigenvalues of a
matrix of (much smaller!) size $k+2$:

\begin{equation}
{\cal{M}}_k=
\begin{pmatrix}
  f''& \dot{f'}_1& \ldots&
  \dot{f'}_{k+1}\\
  \frac{-2}{(1-m_{1})} \dot{f'}_1& \frac{-2}{(1-m_{1})}
  \Ddot{f}_{11}&&
  \frac{-2}{(1-m_{1})} \Ddot{f}_{1(k+1)}\\
  \frac{-2}{(m_1-m_{2})} \dot{f'}_2& \frac{-2}{(m_1-m_{2})}
  \Ddot{f}_{21}&&
  \frac{-2}{(m_1-m_{2})} \Ddot{f}_{2(k+1)}\\
  \vdots&&\ddots&\vdots\\
  \frac{-2}{(m_{(k-1)}-m_{k})} \dot{f'}_{k}&
  \frac{-2}{(m_{(k-1)}-m_{k})} \Ddot{f}_{k1}& \ldots&
  \frac{-2}{(m_{(k-1)}-m_{k})} \Ddot{f}_{k(k+1)}\\
  \frac{-2}{m_k} \dot{f'}_{(k+1)}& \frac{-2}{m_k} \Ddot{f}_{(k+1)1}&
  \ldots&
  \frac{-2}{m_k} \Ddot{f}_{(k+1)(k+1)}\\
\end{pmatrix}
\end{equation}
Here, we have introduced the shorthand notations
\begin{equation}
f''=\frac{\partial^2 f}{\partial \qs^2}, \quad
\dot{f'}_i = \frac{\partial}{\partial q_i}\frac{\partial f}{\partial \qs}, \quad \mbox{and} \quad
\ddot{f}_{ij} = \frac{\partial}{\partial q_i}\frac{\partial f}{\partial q_j}
\end{equation}

\begin{figure}
  \centerline{\input{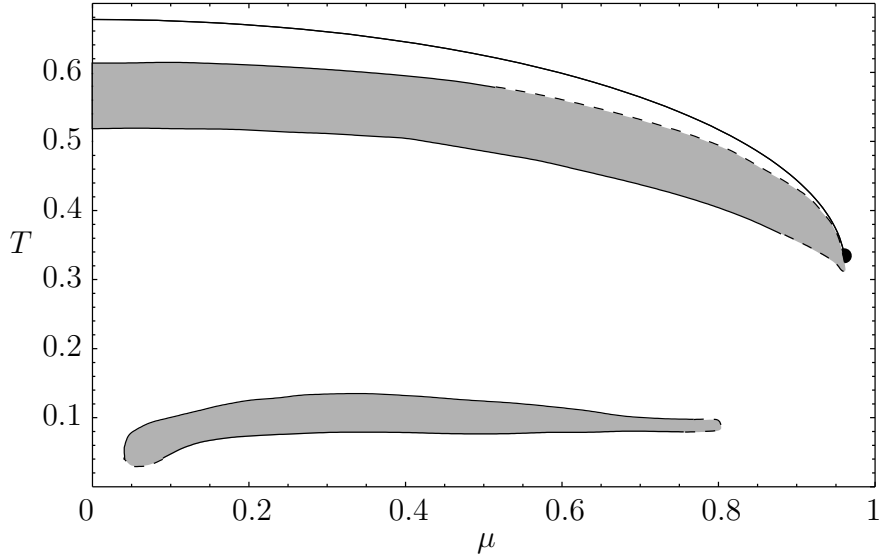}} \vspace{0.3cm}
\caption{The gray regions show complex AT eigenvalues in
  one--step replica symmetry breaking. Dashed boundaries indicate
  difficulties in the numerical algorithm. Here, the true boundary may
  be slightly shifted.}
\label{fig:stability1rsb}
\end{figure}

We have carried out the above calculations numerically for 1RSB with
the order parameters presented in section \ref{sec:nonzeroT}.  Regions
with complex replica--diagonal eigenvalues are shown in Fig.
\ref{fig:stability1rsb}. Now there appear two of them.  Since $f$ and
its derivatives are real, obviously complex eigenvalues always have to
appear in pairs conjugate to each other.  Labelling the three
eigenvalues at 1RSB in proper order, we find that in the
high--temperature region of Fig. \ref{fig:stability1rsb} $\lambda_1$
and $\lambda_2$ turn complex, while in the low--temperature region
$\lambda_2$ and $\lambda_3$ have nonzero imaginary parts. Therefore,
the two regions can be clearly distinguished.  The high--temperature
region has approximately the same shape as the one from RS
calculations. Its boundary at $\mu=0$ is given by $T=0.52$ and
$T=0.61$. This fact is a strong indication that the effect of RSB is
small in the corresponding temperature range and that the occurance of
complex AT eigenvalues continues for arbitrary RSB. Complex
eigenvalues thus seem to be a consequence of the replica limit.  Their
interpretation as far as stability is concerned remains an open
question.

\section{Conclusion}
Using the equivalence of the thermodynamic properties of the
Ghatak--Sherrington model and the fermionic quantum Ising spin glass,
we simultaneously solved these two models in the first step of the
replica symmetric approximation combining analytical and numerical
methods. We also obtained the $T=0$ limit of these solutions. As
expected, the effects of RSB are comparatively large in general, while
the order parameters of charge are almost unchanged by RSB in certain
regions of the ordered phase.  Analyzing the replica--diagonal AT
eigenvalues, we found that the puzzling complex stability eigenvalues,
previously found in RS calculations, also appear in one step RSB. They
even cover a larger part of the phase diagram. On this basis, we
formulated the conjecture that complex eigenvalues of the stability
matrix appear at any order of RSB.

This work was supported by the DFG project Op28/5-1, by the SFB410 and
by the Villigst foundation.

\end{document}